\documentclass{PoS}
\usepackage{graphicx}
\usepackage{amsmath}
\usepackage{subfig}
\usepackage{xspace}
\usepackage{cite}
\usepackage[font=small]{caption}

\newcommand{\hthatprime}{\hat{H}^\prime_T}
\newcommand{\alphaS}{\alpha_\mathrm{s}}
\newcommand{\Higgs}{\ensuremath{\mathrm{H}}\xspace}
\newcommand{\Hj}{\Higgs\!+\! 1\! jet\xspace}

\newcommand{\Hjj}{\Higgs\!+\! 2\! jets\xspace}
\newcommand{\Hjjj}{\Higgs\!+\! 3\! jets\xspace}

\newcommand{\Hnj}{\Higgs\!+\! $n$\! jets\xspace}
\newcommand{\GeV}{\ensuremath{\mathrm{GeV}}\xspace}
\newcommand{\TeV}{\ensuremath{\mathrm{TeV}}\xspace}

\title{Higgs + Multi-jets in Gluon Fusion}

\ShortTitle{Higgs + Multi-jets in Gluon Fusion}

\author{\speaker{Nicolas Greiner}\\
         Physik-Institut, Universit\"at Z\"urich, Wintherturerstrasse 190, CH-8057 Z\"urich, Switzerland\\
         E-mail: \email{greiner@physik.uzh.ch}}

\author{Stefan H\"oche\\
        SLAC National Accelerator Laboratory, Menlo Park, CA 94025, USA\\
        E-mail: \email{shoeche@slac.stanford.edu}}
\author{Gionata Luisoni\\
        Theoretical Physics Department, CERN, Geneva, Switzerland \\
        E-mail: \email{gionata.luisoni@cern.ch}}        
\author{Marek Sch\"onherr\\
        Physik-Institut, Universit\"at Z\"urich, Wintherturerstrasse 190, CH-8057 Z\"urich, Switzerland\\
        E-mail: \email{marek.schoenherr@physik.uzh.ch}} 
\author{Jan-Christopher Winter\\
        Department of Physics and Astronomy, Michigan State University, East Lansing, MI 48824, U.S.A. \\
        E-mail: \email{jwinter@cern.ch}}
\author{Valery Yundin\\
        Max-Planck-Institut f\"ur Physik, F\"ohringer Ring 6, D-80805 M\"unchen, Germany\\
        E-mail: \email{yundin@mpp.mpg.de}}                        

 \abstract{We present a detailed phenomenological analysis of the production of a Standard Model Higgs boson
in association with up to three jets. The Higgs is produced via gluon fusion, which is an irreducible
background to the vector boson fusion mechanism. We calculate the next-to-leading order corrections in QCD
in the limit of an infinitely heavy top quark. Numerical results are presented for a large variety of
observables, for different selection cuts, and for different choices of the jet tagging scheme.
\begin{flushright}
MSUHEP-160114, SLAC-PUB-16455, ZU-TH-1/16
\end{flushright}
}

 \FullConference{12th International Symposium on Radiative Corrections (Radcor 2015) and LoopFest XIV (Radiative Corrections for the LHC and Future Colliders)\\
 15-19 June, 2015\\
 UCLA Department of Physics \& Astronomy Los Angeles, USA}

\begin{document}

\section{Introduction}
A major step after the discovery of the Higgs particle~\cite{Aad:2012tfa,Chatrchyan:2012ufa} is the precise determination of its nature.
This includes its couplings to fermions and bosons. As different models lead to various deviations from the Standard Model prediction,
a precise prediction for the Standard Model contribution is an essential step to the revelation of the underlying mechanism of
electroweak symmetry breaking. During Run~II at the LHC, the vector boson fusion mechanism will play a leading role. 
In this production mode, a Higgs boson is created by annihilation 
of virtual $W$ or $Z$ bosons, radiated off the initial-state (anti-)quarks in a $t$-channel scattering process with no 
color exchange at leading order~\cite{Cahn:1983ip,Kane:1984bb}. It allows for direct access to the couplings between the 
Higgs and the electroweak gauge bosons while at the same time providing a clean signature with two forward jets with only
little hadronic energy between these tagging jets.\\
The main production mechanism is however given by the gluon fusion channel where the Higgs is produced out of two initial state gluons 
via a loop of heavy quarks. The gluon fusion mechanism constitutes an interesting process on its own, but it is also an irreducible 
background to the VBF channel. In both cases a reliable theoretical prediction is indispensable.
In this talk we discuss the calculation and the phenomenlogy of the production of a Standard Model Higgs boson in association
up to three jets, as described in detail in Refs.~\cite{Greiner:2015jha,Cullen:2013saa,vanDeurzen:2013rv}. 
The calculation takes into account next-to-leading order QCD corrections and it is carried out in the 
limit of an infinitely heavy top quark. We discuss two different sets of cuts, namely a set of basic cuts that are suitable for the 
gluon fusion contribution and a more restrictive set of cuts which is more suited for VBF analyses. In addition we discuss different
schemes for defining the tagging jets.

\section{Calculational setup}
We perform the calculation of the NLO corrections by using the automated tools \textsc{GoSam}
\cite{Cullen:2011ac,Cullen:2014yla} for the generation
of the virtual amplitudes, and \textsc{Sherpa} \cite{Gleisberg:2008ta} for the tree-level like amplitudes, subtraction terms and phase space
integration. The two are linked
using the Binoth Les Houches Accord
\cite{Binoth:2010xt,Alioli:2013nda}, a standard interface for event and
parameter passing between one-loop programs and Monte Carlo
generators.\\
The one-loop program \textsc{GoSam} is based on an algebraic generation of
$d$-dimensional integrands using a Feynman diagrammatic approach,
employing \textsc{QGraf}~\cite{Nogueira:1991ex} and
\textsc{Form}~\cite{Vermaseren:2000nd,Kuipers:2012rf} for the diagram
generation, and \textsc{Spinney}~\cite{Cullen:2010jv},
{\textsc{Haggies}}~\cite{Reiter:2009ts} and \textsc{Form} to write an
optimized Fortran output. For the reduction of the tensor integrals,
we used \textsc{Ninja}~\cite{Mastrolia:2012bu,vanDeurzen:2013saa,Peraro:2014cba},
which is an automated package carrying out the integrand reduction via
Laurent expansion. Alternatively, one can use other reduction
techniques such as the standard OPP
method~\cite{Ossola:2006us,Mastrolia:2008jb,Ossola:2008xq} as
implemented in \textsc{Samurai}~\cite{Mastrolia:2010nb} or methods of
tensor integral reduction as implemented in
\textsc{Golem95}~\cite{Heinrich:2010ax,Binoth:2008uq,Cullen:2011kv}.
The resulting scalar integrals are evaluated using
\textsc{OneLoop}~\cite{vanHameren:2010cp}. More details on the calculation and reduction of multi-loop integrands
have been presented at this workshop by Giovanni Ossola in Ref.~\cite{giovanni}.\\
The calculation of tree-level matrix elements real emission contribution as well as the subtraction terms in the 
Catani-Seymour approach~\cite{Catani:1996vz} has been done within \textsc{Sherpa}~\cite{Gleisberg:2008ta}
using the matrix element generator \textsc{Comix}~\cite{Gleisberg:2008fv,Hoeche:2014xx}. We have validated the 
results obtained by \textsc{Comix} with the results we have obtained in Ref.~\cite{Cullen:2013saa}, where we
have used a combination of \textsc{MadGraph~4}
\cite{Stelzer:1994ta,Alwall:2007st}, \textsc{MadDipole}
\cite{Frederix:2008hu,Frederix:2010cj} and \textsc{MadEvent}
\cite{Maltoni:2002qb} for the calculation of real emission matrix
elements, subtraction terms and the phase space integration of the
real emission contribution.

\subsection{Cuts and parameter settings}
\label{sec:cuts}
We have produced results for two center of mass energies at 8 and 13 TeV. In both cases
we have applied two sets of cuts, one baseline set which is based on a minimal set of cuts
to render the cross section finite, and a more restrictive set of cuts which is typically 
used in the context of VBF searches. In both cases jets
are clustered using the anti-$k_T$
algorithm~\cite{Cacciari:2005hq,Cacciari:2008gp} as implemented in the
\textsc{FastJet} package~\cite{Cacciari:2011ma}. If not specified
explicitly, the jet radius and PDF set have been set to $R=0.4$ and
CT10nlo~\cite{Lai:2010vv}, respectively. The baseline set consists of the following cuts:
\begin{equation}
 p_T\;>\;30~\text{GeV}~,\qquad |\eta|\;<\;4.4~.
 \label{cuts:basic}
\end{equation}
In the VBF case two additional cuts have been imposed, given by
\begin{equation}
  m_{j_1 j_2}\;>\;400~\text{GeV}~,\qquad \left|\Delta y_{j_1,\,j_2}\right|\;>\;2.8~.
  \label{cuts:vbf}
\end{equation}
Here the two jets, $j_1$ and $j_2$, denote the tagging jets. Their selection is not unique and we
study two different schemes, one where the two tagging jets are the two jets with the highest $p_T$
($p_T$-tagging), and one where the most forward and most backward jet (in rapidity) yield 
the two tagging jets ($y$-tagging).\\
The central scale for the variation of remormalization and factorization scale is chosen to be
\begin{equation}
\mu_{F}=\mu_{R}\;\equiv\;\frac{\hthatprime}{2}\;=\;
\frac{1}{2}\left(\sqrt{m_{H}^{2}+p_{T,H}^{2}}+\sum_{i}|p_{T,i}|\right)\;.
\end{equation}
However it is not obvious whether this dynamical scale is also a good choice to be used for 
the gluon-gluon-Higgs coupling. One might argue that the Higgs mass is the appropriate scale 
there. Therefore we consider three different scale choices, defined as
\begin{subequations}
\label{scales}
\begin{flalign}
\label{scales:A}
\textrm{A}:&\qquad\alphaS\biggl(x\cdot\frac{\hthatprime}{2}\biggr)^3\;
            \alphaS\left(x\cdot m_{H}\right)^2~,\\[1mm]
\label{scales:B}
\textrm{B}:&\qquad\alphaS\biggl(x\cdot\frac{\hthatprime}{2}\biggr)^5~,\\[1mm]
\label{scales:C}
\textrm{C}:&\qquad\alphaS\left(x\cdot m_{H}\right)^5~.
\end{flalign}
\end{subequations}
The presence of the factor $x$ indicates that this scale is varied in
the range $x\in[0.5\,\ldots\,2]$.

\section{Numerical results for gluon fusion setup}
\begin{figure}[t!]
  \centering
  \includegraphics[width=0.49\textwidth]{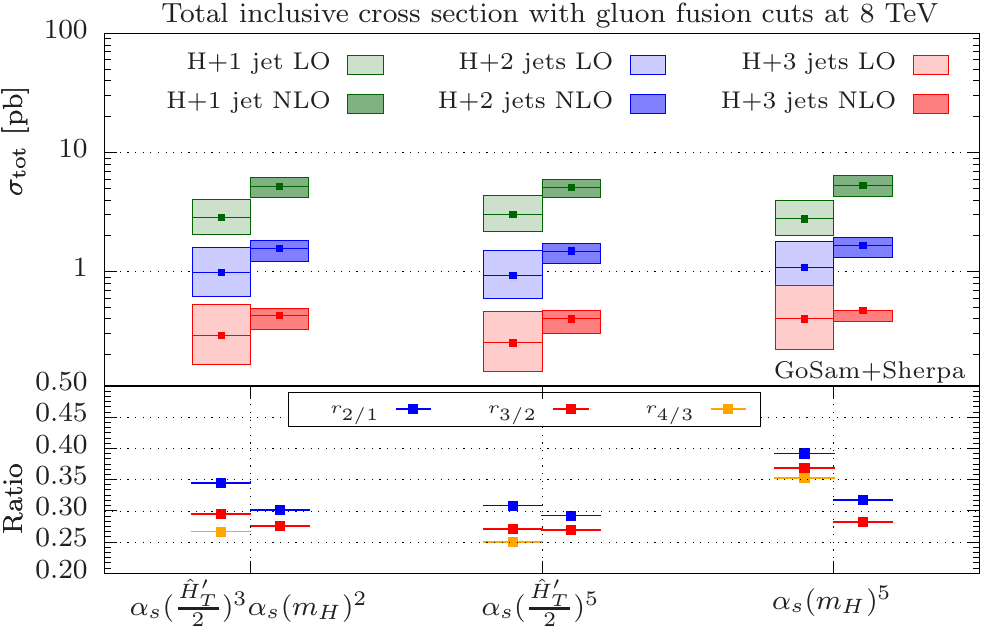}
  \hfill
  \includegraphics[width=0.49\textwidth]{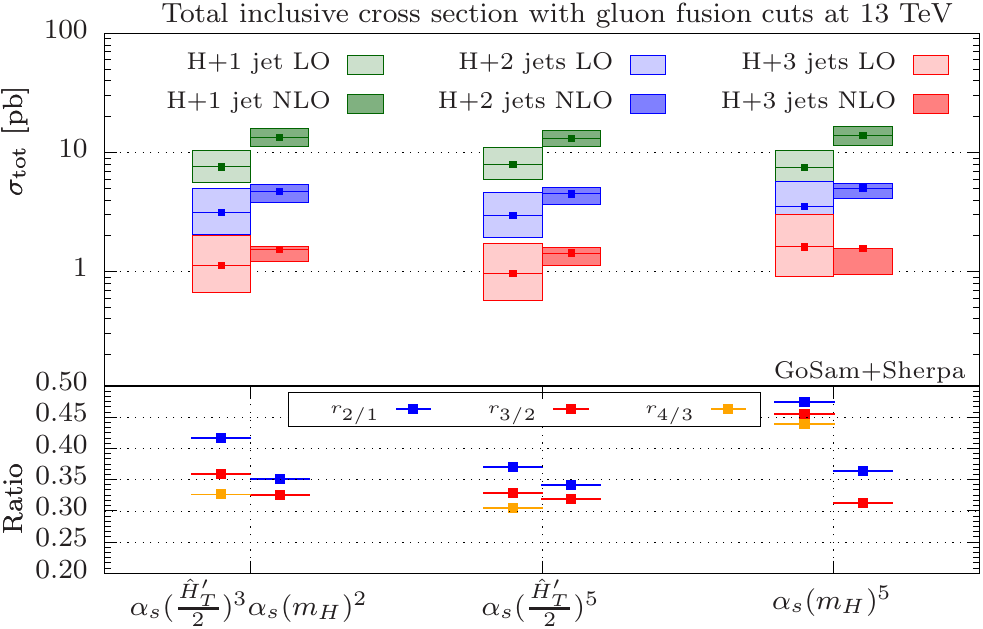}
  \caption{\label{fig:Xsec}%
    Total cross sections at LO (left side of each column) and NLO
    (right side of each column) for \Hj (green), \Hjj (blue) and \Hjjj
    (red) production using the three different scale choices as explained in text. In the
    lower part of the plots, the ratios $r_{2/1}$ (blue), $r_{3/2}$
    (red) and $r_{4/3}$ (orange) are shown. Results have been obtained
    for  8~\TeV and 13~\TeV (left and right plot respectively).}
\end{figure}

\begin{figure}[t!]
  \centering~\hfill
  \includegraphics[width=0.49\textwidth]{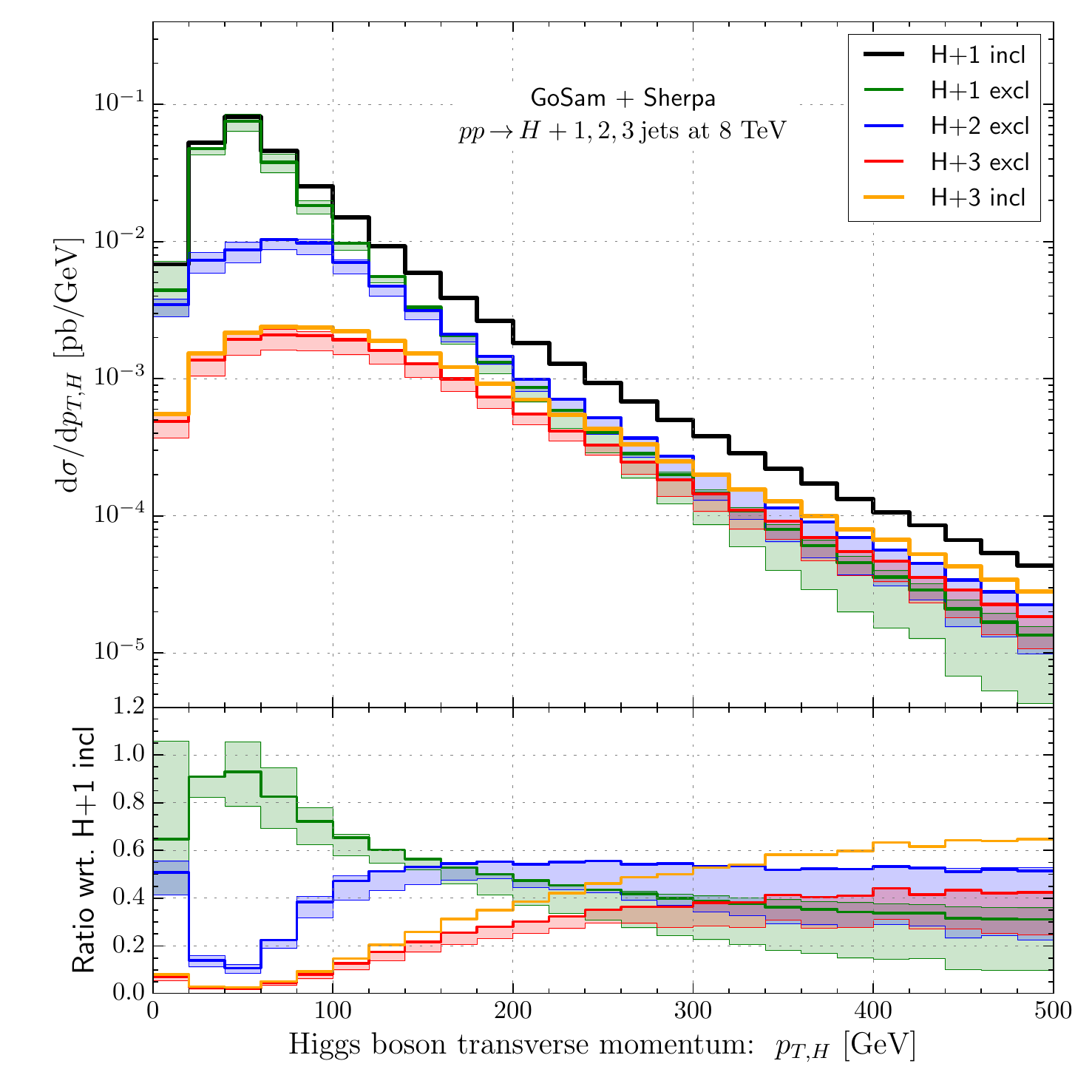}
  \hfill
  \includegraphics[width=0.49\textwidth]{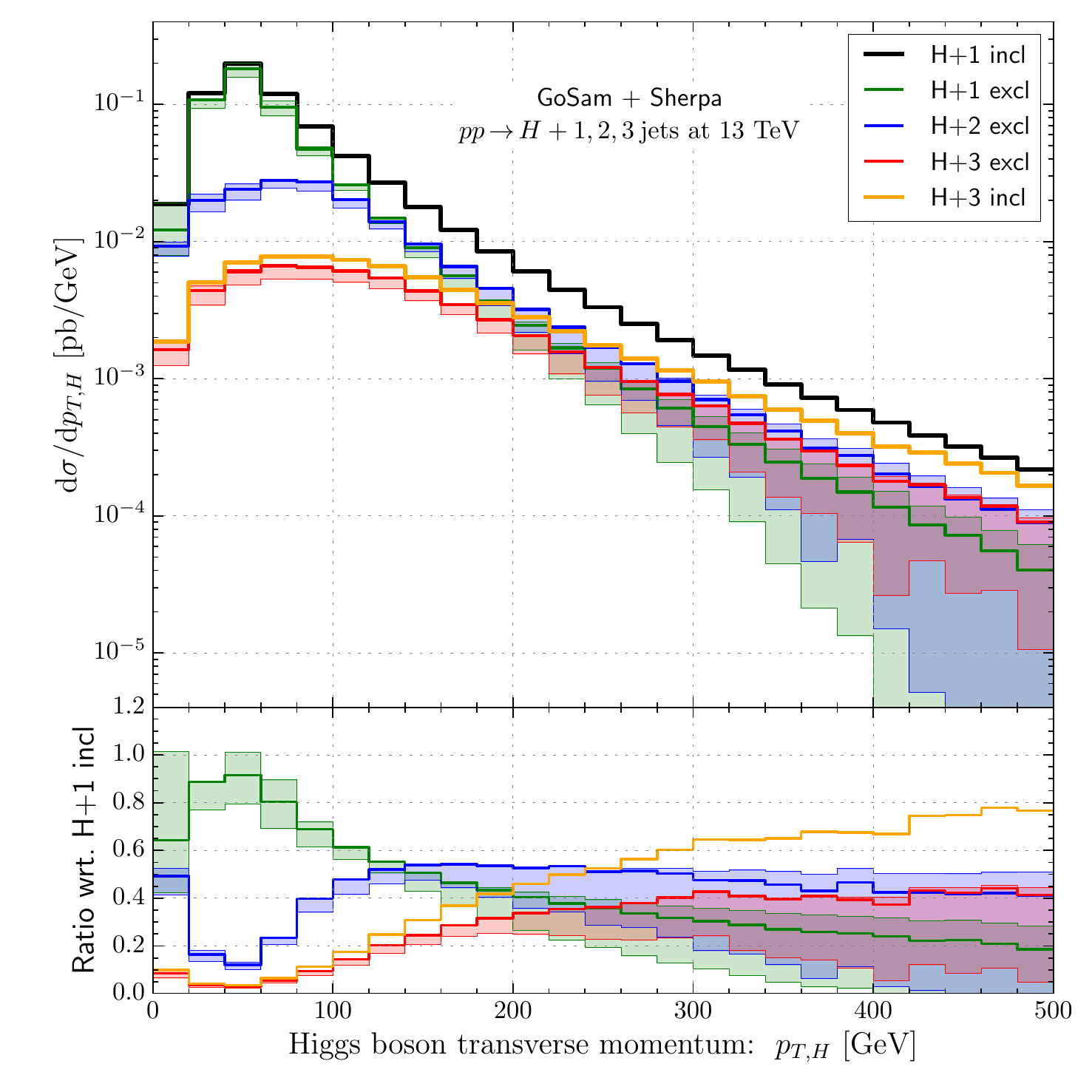}
\caption{\label{fig:xnlo-ratio-HTprime} Transverse momentum distribution of the Higgs for the different jet multiplicities. 
L.h.s. shows the result for 
8 TeV, r.h.s. shows the distribution for 13 TeV.}
\end{figure}

We start the discussion of the numerical results with the basic gluon fusion cuts.
One of the most important observables is the total cross section. In Figure~\ref{fig:Xsec} 
we show the total cross sections
for both leading order and next-to-leading order for the processes \Hj, \Hjj and \Hjjj at
$E_\mathrm{cm}=8$~\TeV (left plot) and $E_\mathrm{cm}=13$~\TeV (right plot).
The result have been obtained for the three scale choices described above. On the level 
of total cross sections one only observes a mild dependence on the scale choice, in particular 
for the NLO results. For the fixed scale one observes an enhancement of the LO ratios.
An increase of the center of mass energy from $8$ to $13$ TeV also only has a small influence
on the overall pattern.
\begin{figure}[t!]
  \centering
  \subfloat[Ratio\label{fig:H3_higgs_scales:scaleRatio}]{\includegraphics[width=0.49\textwidth,height=74mm]{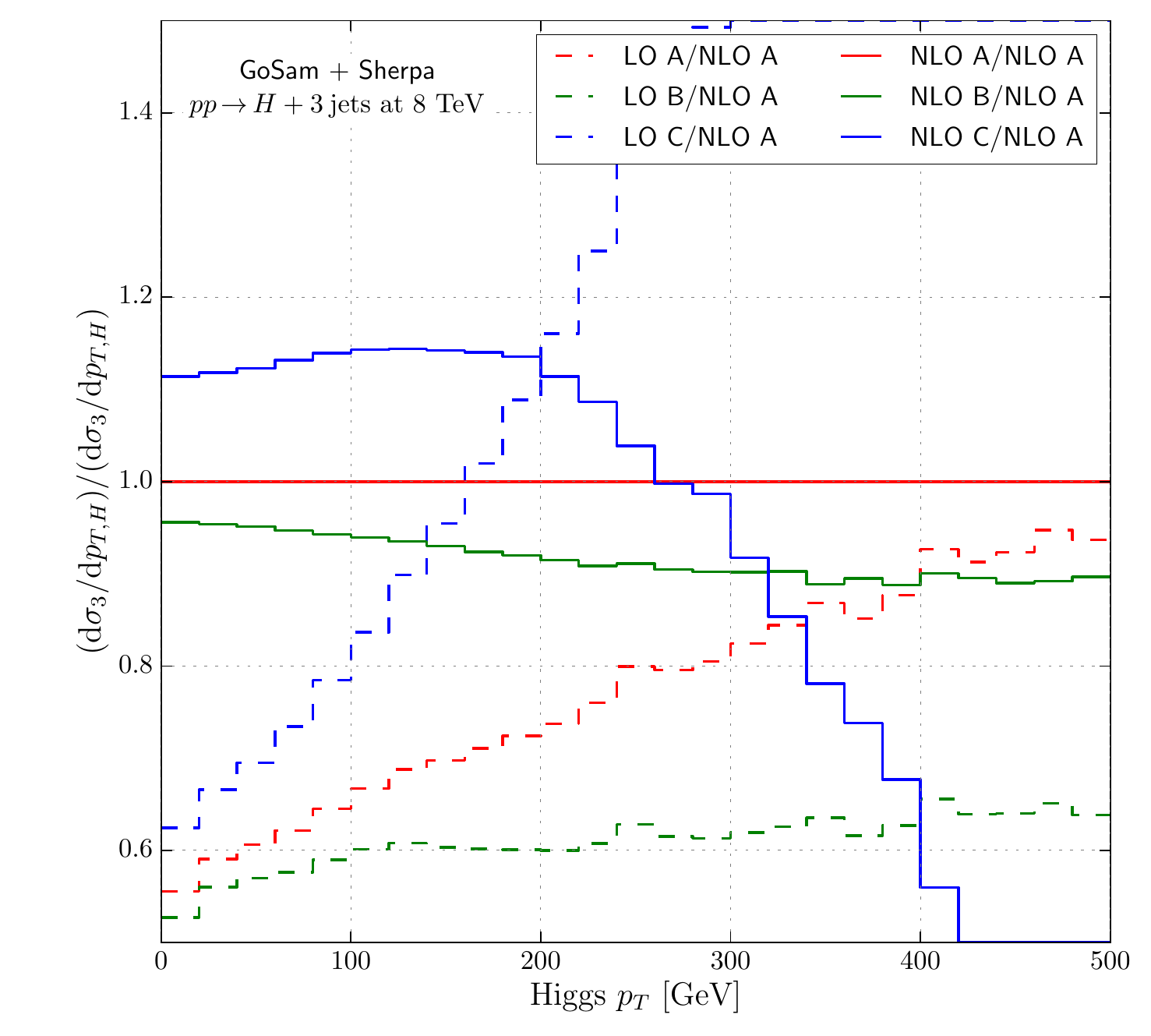}}
   \hfill
   \subfloat[Scale choice A \label{fig:H3_higgs_scales:scaleA}]{\includegraphics[width=0.49\textwidth]{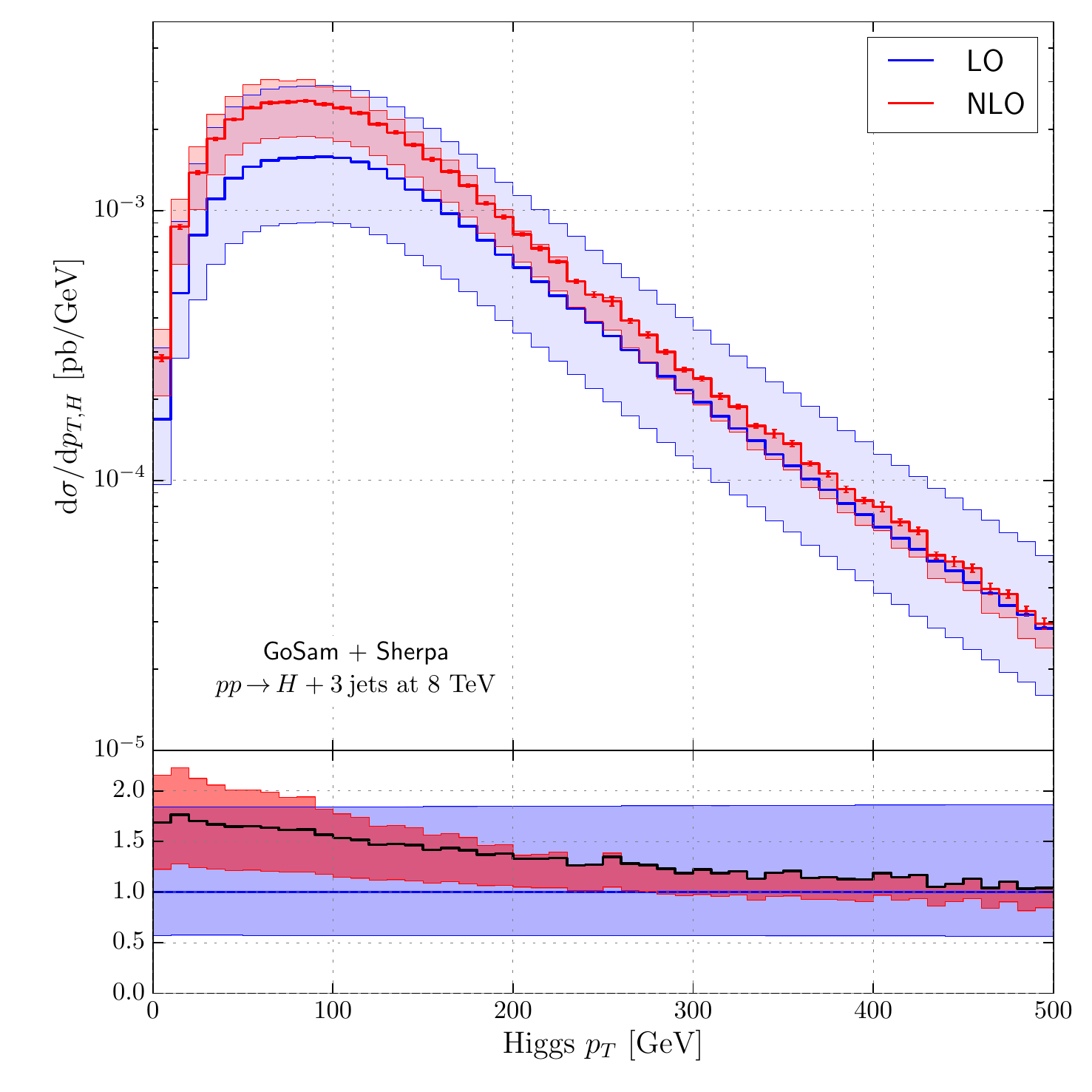}}
  \\
  \centering
  \subfloat[Scale choice B \label{fig:H3_higgs_scales:scaleB}]{\includegraphics[width=0.49\textwidth]{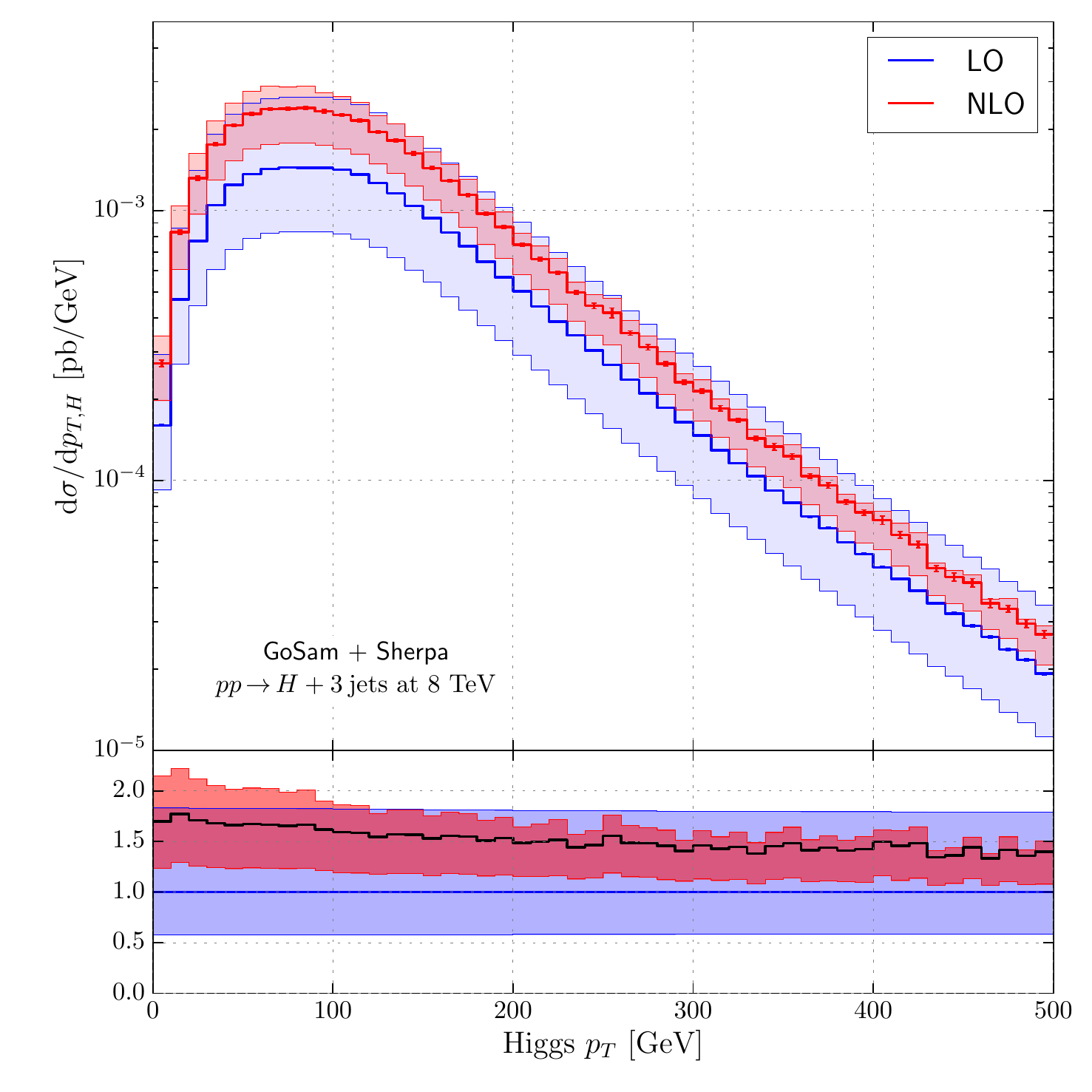}}
  \hfill
  \subfloat[Scale choice C \label{fig:H3_higgs_scales:scaleC}]{\includegraphics[width=0.49\textwidth]{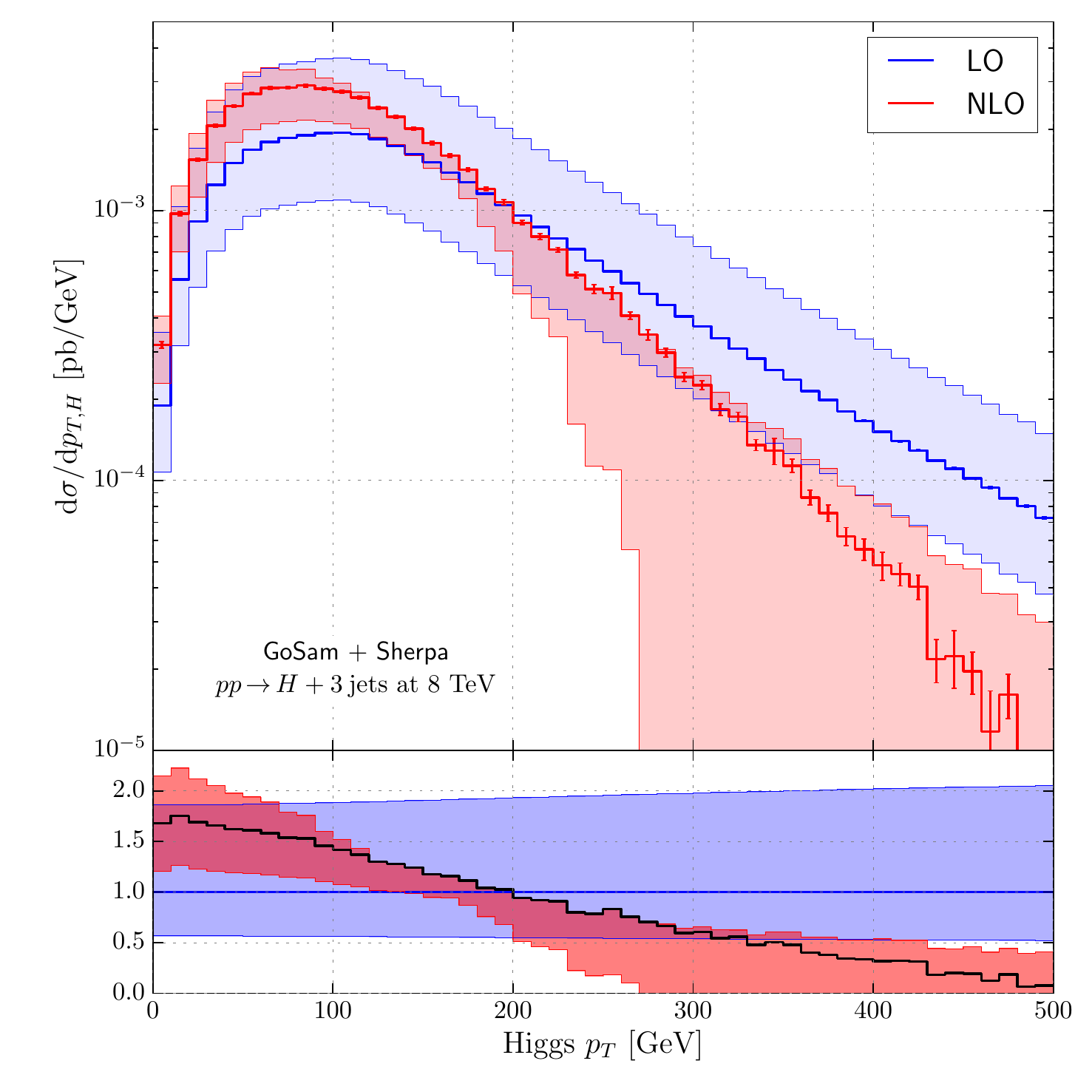}}
   \caption{\label{fig:H3_higgs_scales}%
     The $p_T$ distribution of the Higgs boson in \Hjjj production at
     the 8~\TeV LHC presented for the three scales A, B and C . The subplot
     \protect\ref{fig:H3_higgs_scales:scaleRatio} shows the same central
    predictions normalized to the NLO result for scale A. Each ratio
    plot depicts the respective differential $K$-factors and their
    envelopes obtained from scale variations at LO and NLO.}
\end{figure}
The situation is different when one looks at differential distributions. In Figure~\ref{fig:H3_higgs_scales} we show the 
transverse momentum distribution for the Higgs boson for the \Hjjj process at a center of mass energy of 8 TeV.
The subplots show the distribution for the three different scale choices A, B and C,
while Fig.~\ref{fig:H3_higgs_scales:scaleRatio}
shows the results for the different scales normalized to the NLO
result for scale A. The advantage of scale B is the flatness of the
$K$-factor over the entire $p_T$ range. This supports our choice to
make scale B the default scale. For the lower $p_T$ region up to
$\sim250$~\GeV, scale C seems to be a sensible choice as
well. However, it completely breaks down for higher $p_T$, and the
$K$-factor can even become negative.
Further support for using scale B as the default choice comes from looking at the $p_T$ distribution of the
'wimpiest' jet for each multiplicity. This means looking at the first jet for \Hj, at the second jet for \Hjj,
and at the third jet for \Hjjj. 
\begin{figure}[t!]
  \centering
  \parbox[t]{\textwidth}{
  \includegraphics[width=0.49\textwidth]{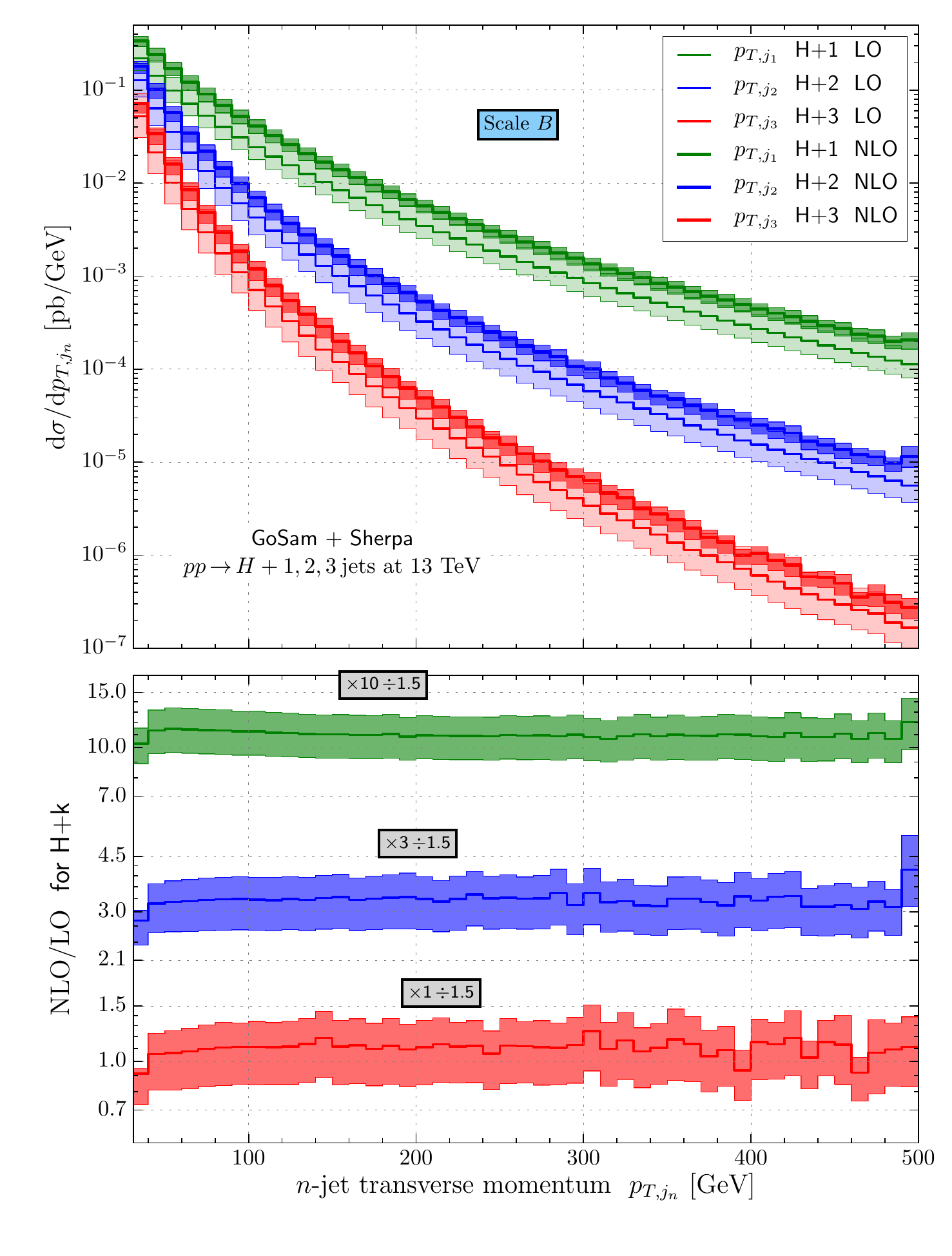}
  \hfill
  \includegraphics[width=0.49\textwidth]{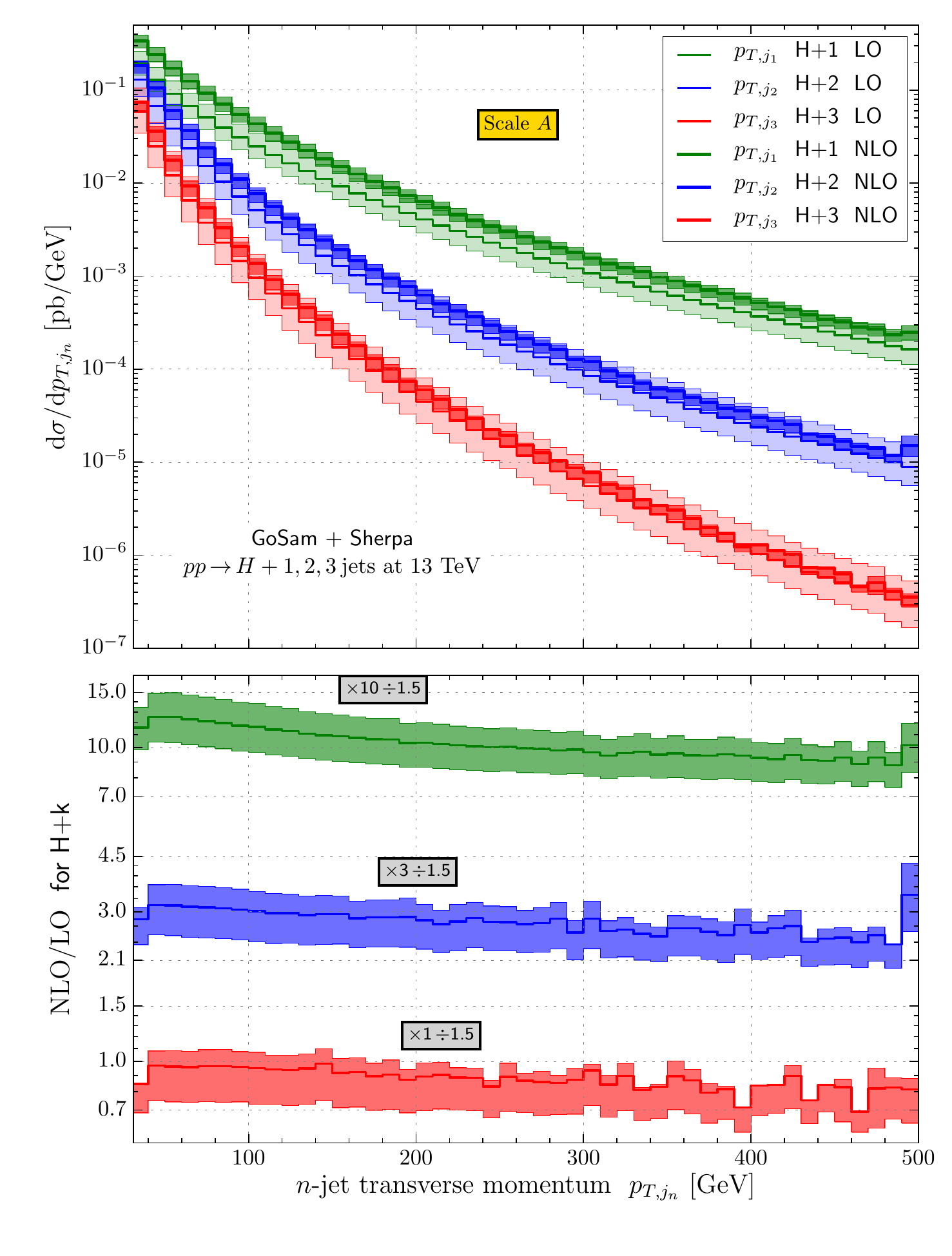}
  }
   \caption{\label{fig:subleadingjetpt}%
     Transverse momentum distribution of the `wimpiest' jet in \Hnj
    production at the LHC. Using $p_T$ ordering the first, second and
    third leading jet are shown in \Hj, \Hjj and \Hjjj at 13~\TeV,
    respectively; on the left with the default scale choice B, on the
    right with the scale choice A. }
\end{figure}
This is illustrated in Figure~\ref{fig:subleadingjetpt}. The left hand side shows the $p_T$ of the three jets 
for scale choice B, the right hand side shows scale choice A. The lower part of the plots shows the ratio of NLO versus
LO for each jet. For better visibility the ratios are multiplied with a certain factor to avoid overlap of the bands 
of the different jets. As can be seen from the ratio plots, the purely dynamical scale choice B leads to flat K-factors,
whereas scale choice A shows a decrease of the K-factor with increasing transverse momentum.\\ \\
Another interesting question is how observables that are defined independent of a certain jet multiplicity, like 
the transverse momentum of the Higgs, changes under the presence of additional QCD radiation.
This is shown in Figure~\ref{fig:xnlo-ratio-HTprime}.
The upper plots show the NLO
distributions for one, two and three jets (which we have obtained from
the one-jet, two-jet and three-jet NLO calculations, respectively).
Unless stated otherwise, the jet multiplicity is exclusive, labelled
by `excl', i.e.~a veto on any additional jet activity is in place. The
1-jet and the 3-jet processes are shown twice, once for the exclusive
case, and once for the inclusive case, labelled by 'incl'. The lower
subpanels show each contribution normalized to the inclusive
prediction of the core process, i.e.~the most inclusive one, here
given by the \Hj process. The plots in the middle and lower panel are
constructed following the same principle but using the NLO core
process of increased jet multiplicity, namely \Hjj and \Hjjj,
respectively. The middle row of Fig.~\ref{fig:xnlo-ratio-HTprime}
hence depicts the same situation but without accounting for the \Hj
process; and, for the lower row, there are only two distributions left
to show, the ones for the exclusive as well as the inclusive \Hjjj
process. One can see that the low energy region is dominated by the exclusive 
\Hj contribution. The \Hjj contribution is negligible in that region, however 
starts to dominate already in a region above approx. 200 \GeV. Going further 
up in the spectrum increases the \Hjjj contribution which will eventually dominate 
the spectrum. In other words, at high enough energies, to have one jet more comes with 
the same or even higher probability. This has to be kept in mind when comparing an inclusive 
measurement with a fixed order calculation for a given number of jets. For
low multiplicities the description becomes inaccurate already at relatively low 
energies of around 200 \GeV. A theoretical prediction that is based on a merged result of 
different multiplicities will yield a better description of the data.

\section{Phenomenology with vector boson fusion cuts}
Gluon fusion is an irreducible background to the VBF channel, the
challenging task for theory is therefore to provide a precise prediction of its
rate compared to the signal. In this section we discuss the results obtained from the 
gluon fusion contribution under the presence of the additional VBF cuts described in
section \ref{sec:cuts}. Again we start the discussion with the total cross section, for 
the VBF selection we now also consider the differences between the two tagging schemes described 
above, $p_T$-tagging and $y$-tagging.
\begin{figure}[t!]
  \centering
  \includegraphics[width=0.49\textwidth]{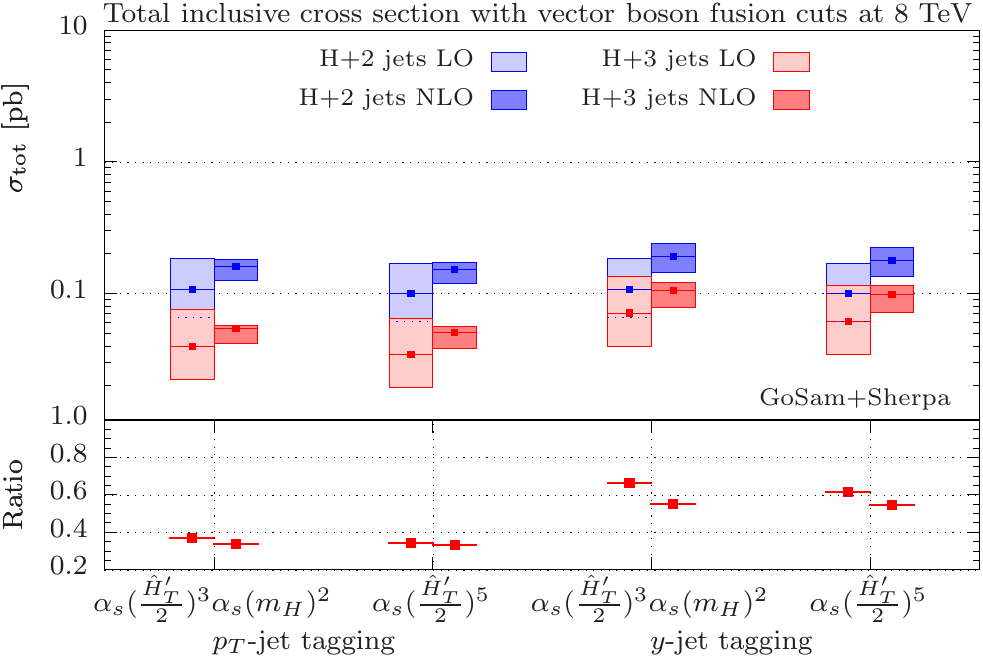}
  \hfill
  \includegraphics[width=0.49\textwidth]{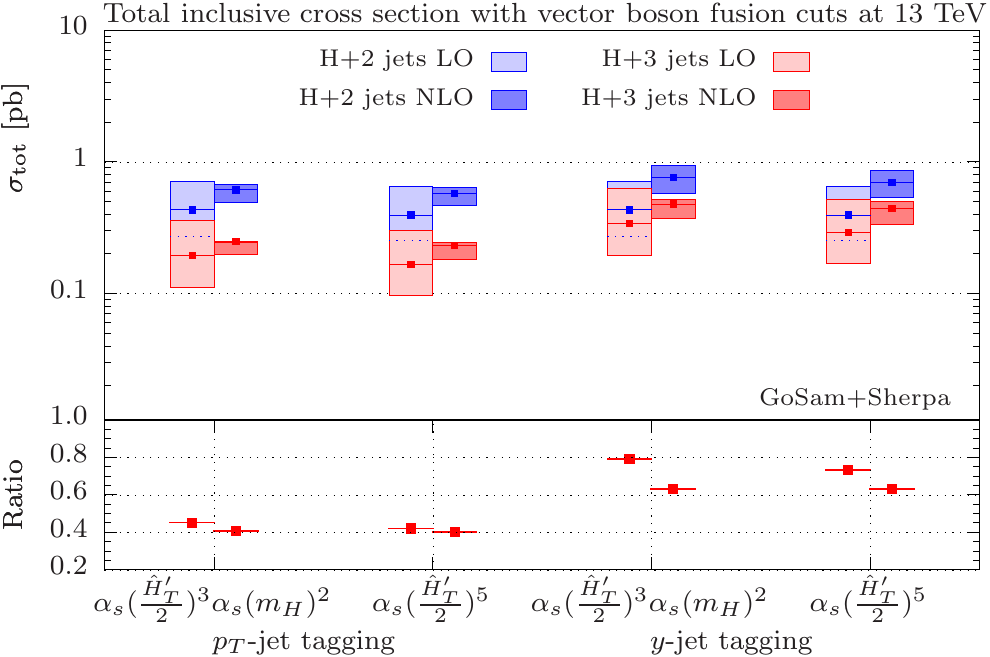}
  \caption{\label{fig:VBF_xsec}%
    Total cross sections at LO  and NLO  for \Hjj (blue) and \Hjjj (red)
    using VBF kinematical cuts and two different tagging jet
    definitions. Results are shown for the two scale choices A and B, as
    well as the two energies of 8~\TeV (left plot) and 13~\TeV
    (right plot). The lower part of each plot depicts the inclusive
    cross section ratios $r_{3/2}$ for the different scales and tag
    jet approaches.}
\end{figure}
The total cross section for the different energies, scales and tagging schemes is shown in Figure~\ref{fig:VBF_xsec}.
Having ruled out the fixed scale as a sensible choice we only show the result for the two scales A and B.
Also in the case of the VBF selection, the differences between the scale choices are rather small. The tagging 
scheme has a much bigger impact. The $y$-tagging increases the ratios which means that it increases the fraction
of the processes with higher multiplicity.
\begin{figure}[t!]
  \centering
  \includegraphics[width=0.49\textwidth]{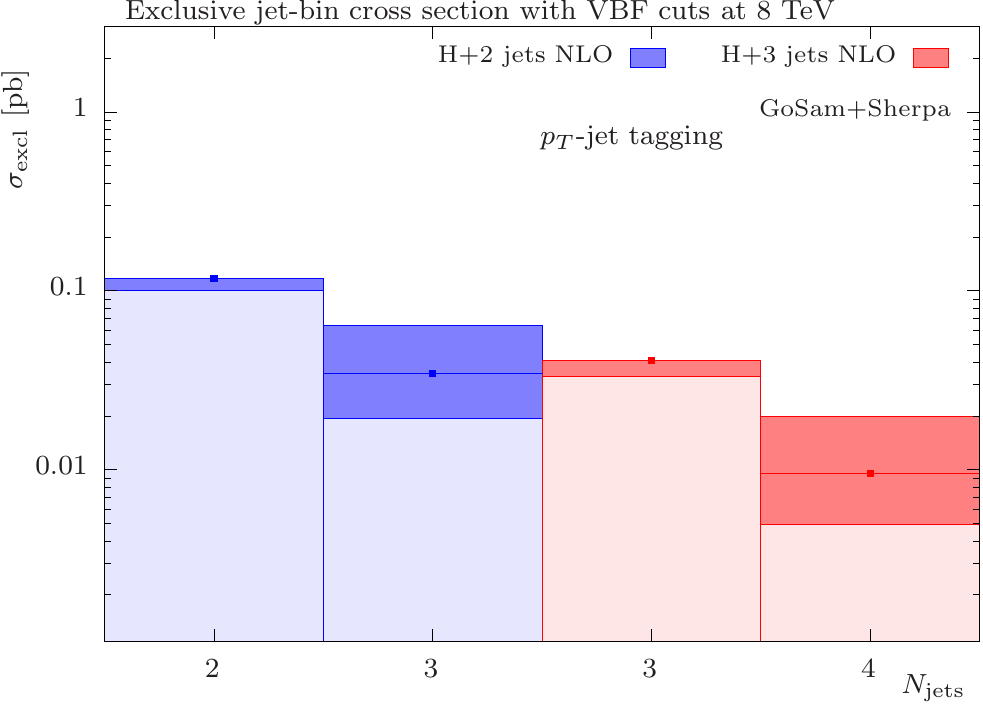}
  \hfill
  \includegraphics[width=0.49\textwidth]{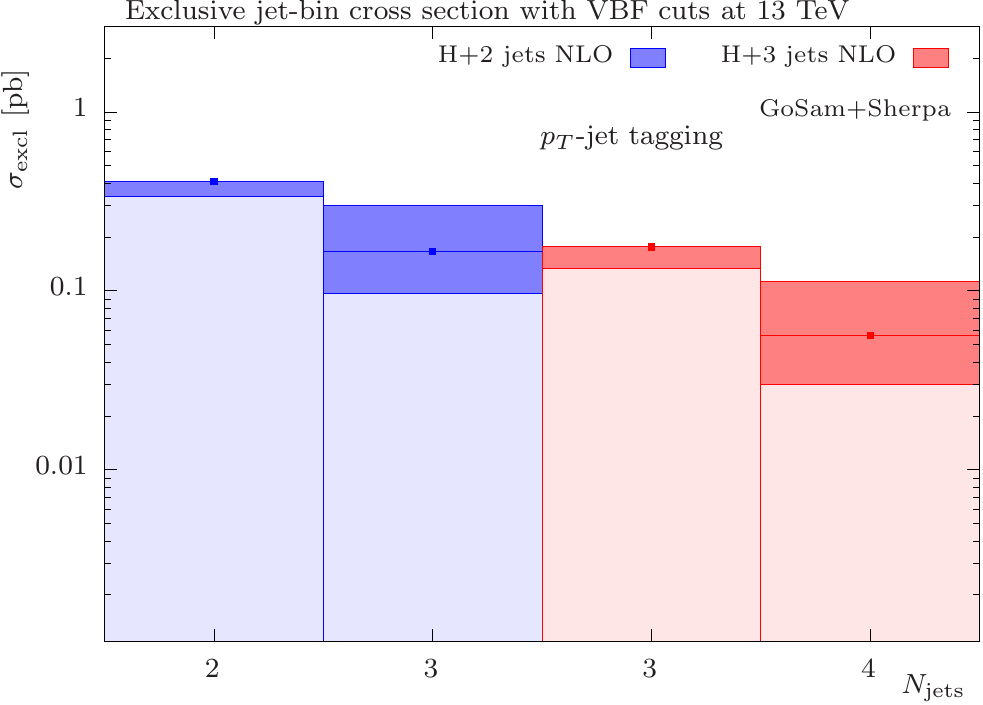}
  \\[4mm]
  \includegraphics[width=0.49\textwidth]{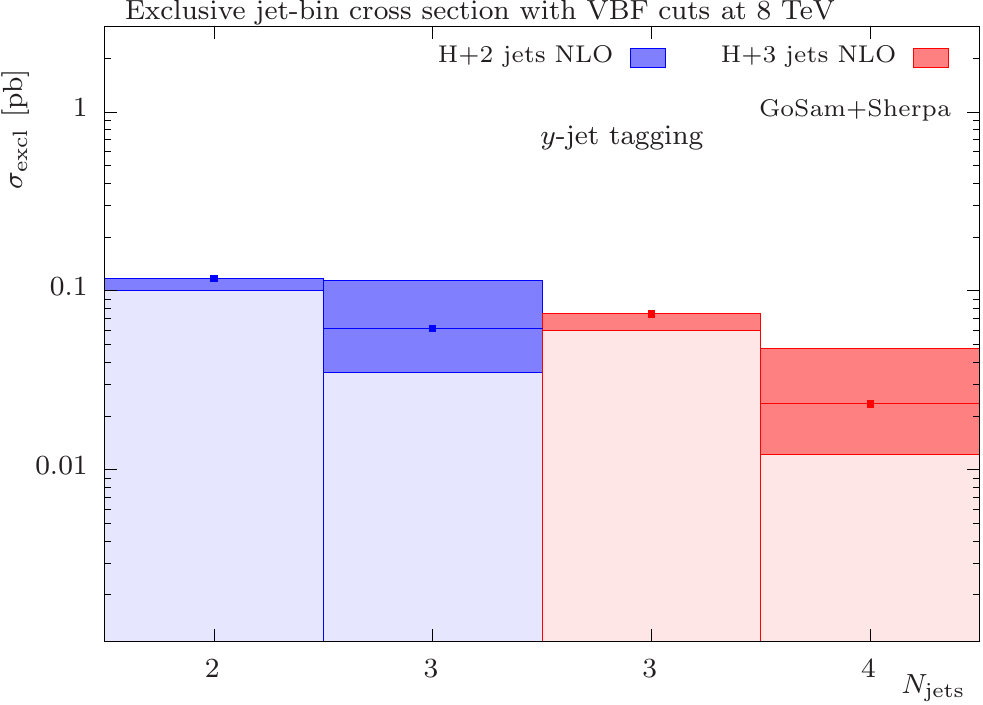}
  \hfill
  \includegraphics[width=0.49\textwidth]{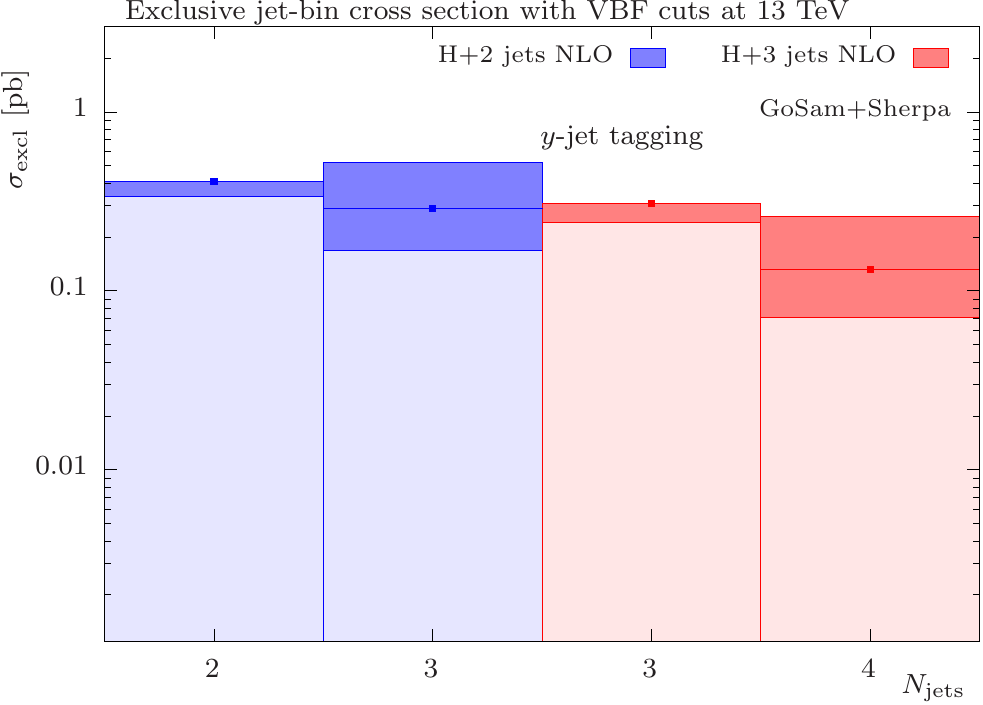}
  \caption{\label{fig:VBF_nj}%
    Exclusive jet cross sections in \Hjj and \Hjjj production at the
    8~\TeV (left) and 13~\TeV (right) LHC after application of typical
    VBF selection constraints (using scale choice B).
    The upper and lower set of plots display the results based on
    different jet-tagging strategies, namely for $p_T$ jet-tagging and
    $y$ jet-tagging, respectively. }
\end{figure}
The impact of the higher jet multiplicities can be understood by looking at the exclusive $n$-jet cross section. This is
shown for \Hjj and \Hjjj in Figure~\ref{fig:VBF_nj}. In the upper row the results are shown using $p_T$ tagging, the lower row
displays the $n$-jet cross section after applying the $y$-tagging scheme. In both cases, but particularly for the $y$-tagging
one sees that the real emission contribution constitutes a substantial fraction of the total cross section. This also means
that a large fraction of the cross section is only described at leading order accuracy. This stresses the importance of the 
inclusion of NLO results with higher multiplicities into the theoretical prediction.

\section{Conclusions}
In this talk we presented phenomenological results for the production of a Standard Model Higgs boson via 
the gluon fusion mechanism in the heavy top mass limit in association with up to three jets. 
We investigated the role of the scale choice as 
well as the effects for different set of cuts, also allowing to assess the role of the gluon fusion contribution in 
VBF searches. Furthermore we discussed a variety of important observables allowing for a better discrimination between
the gluon fusion and vector boson fusion contribution.
Further improvements could certainly be achieved by providing a merged
NLO result of the different jet multiplicities, but also through the
inclusion of top-quark mass effects as well as the matching of the
\Hjjj NLO result with a parton shower.

\section*{Acknowledgments}
We would like to thank the members of the GoSam collaboration for their help and effort.
This work was supported by the Swiss National Science Foundation under contracts PZ00P2\_154829 and
PP00P2--128552 and by the US Department of Energy under contract DE-AC02-76SF00515.

\end{document}